\documentstyle[12pt,epsf]{article}
\newlength{\overeqskip}
\newlength{\undereqskip}
\newcommand{\nc}{\newcommand}

\nc{\be}[1]{\begin{equation} \mbox{$\label{#1}$}}
\nc{\bea}[1]{\begin{eqnarray} \mbox{$\label{#1}$}}
\nc{\Label}[1]{\label{#1}}
\nc{\bi}[1]{\bibitem{#1}}

\nc{\ee}{\end{equation}}
\nc{\eea}{\end{eqnarray}}
\nc{\sss}{\scriptscriptstyle}
\nc{\lsim}{\mbox{\raisebox{-.6ex}{~$\stackrel{<}{\sim}$~}}}
\nc{\gsim}{\mbox{\raisebox{-.6ex}{~$\stackrel{<}{\sim}$~}}}
\nc{\nn}{\nonumber}

\def\NBR#1{{\left( #1 \right)}}                  % Large normal brackets
\def\ie{{\em i.e.\ }}

\def\Re{{\rm Re\,}}

\def\calA{{\cal A}}

\begin{document}
%
% -----------------------------------------------------------------
% Title Page
% -----------------------------------------------------------------
%
\begin{titlepage}
\pagestyle{empty}
\baselineskip=21pt
\rightline{NORDITA-99/30 HE}
\rightline{NBI-HE-99-15}
\rightline{hep-ph/9906411}
\vskip 0.8in

\begin{center} {\Large{\bf The semiclassical propagator in field theory}}

\end{center}
\vskip .3in

\begin{center}

Michael Joyce$^1$,  Kimmo Kainulainen$^2$ and Tomislav Prokopec$^3$\\

\vskip .25in

$^1${\it  INFN, Sezione di Roma 1 and University of Rome, \\
          ``La Sapienza," Ple.\ Aldo Moro 2, 00185 Roma, Italy }\\
\vskip .1in
$^2${\it  NORDITA, Blegdamsvej 17, DK-2100 Copenhagen \O, Denmark}\\
\vskip .1in
$^3${\it  Niels Bohr Institute, Blegdamsvej 17, DK-2100, Copenhagen \O , Denmark }\\

\end{center}
\vskip 0.3 in

\centerline{ {\bf Abstract} }
\baselineskip=18pt
\vskip 0.5truecm

\noindent
We consider scalar field theory in a changing background field. As an example
we study the simple case of a spatially varying mass for which we construct
the semiclassical approximation to the propagator. The semiclassical dispersion
relation is obtained by consideration of spectral integrals and agrees with the
WKB result. Further we find that, as a consequence of localization, the
semiclassical approximation necessarily contains quantum correlations in
momentum space.
 \\

\end{titlepage}

\baselineskip=20pt

%
%%%%%%%%%%%%%%%%%%%%%%%%%%%%%%%%%%%%%%%%%%%%%%%%%%%%%%%%%%%%%%%%%%%%%%%%%%%%%%%
%  MAIN TEXT
%%%%%%%%%%%%%%%%%%%%%%%%%%%%%%%%%%%%%%%%%%%%%%%%%%%%%%%%%%%%%%%%%%%%%%%%%%%%%%%
%
\section{Introduction}

The semiclassical (WKB) method has been developed to describe
motion of a quantum particle in a slowly varying background. The
method is a systematic expansion of the wave function in powers of the
Planck constant, or equivalently, in gradients of the background \cite{wkb}.
In field theory, on the other hand, one is usually concerned with coupling
constant expansions, and applicability of the semiclassical method has
appeared more limited. Nevertheless, there are many cases where the
semiclassical method can be useful also in field theory. For example,
it may be adequate to describe a part of the system by a classical field
to which some other quantum field is weakly coupled.
An important physical example of this occurs at the electroweak phase
transition, where the problem is to study a particle propagation in the
presence of a spatially varying, CP-violating Higgs condensate. Interactions
of the plasma in this background are known to lead to baryon production via
the electroweak anomaly. This  is widely thought to be the most attractive
explanation of the matter-antimatter asymmetry of the Universe \cite{revw},
soon to be tested by accelerator experiments.

The scale at which the Higgs condensate varies is determined by the coupling
of the Higgs field to other species in the plasma and by the dynamics of the
phase boundary during the phase transition \cite{mp}. Quantitative studies
have shown that phase boundaries are often thick when compared with a typical 
De Broglie wave length of a particle in the plasma \cite{cm}. Because
the Higgs condensate endows particles with mass, it is then reasonable to
model the condensate by a spatially varying mass. It is known that source
terms in the dynamical equation leading to baryogenesis, due to these mass
terms, appear beyond leading order in gradient expansion \cite{jpt,cjk}.
However, neither particle propagation, nor plasma dynamics has so far
been studied in a controlled approximation scheme beyond leading order in
gradients.  Other applications of the semiclassical method abound,
{\it e.g.}  motion of electrons in the background of spatially
varying potentials, {\it etc.}

In this letter we develop an approximation in powers of gradients for the
propagator of a scalar field with a spatially varying mass term $m^2 \!
=\! m^2(t,\vec x)$ in a theory described by the lagrangian
\be{Lagrange}
   {\cal L} =
    \NBR{\partial_\mu \phi}^{\dagger} \NBR{\partial^\mu \phi}
    - m^2 \phi^\dagger \phi + {\cal L}_{\rm int}\,,
\ee
where ${\cal L}_{\rm int}$ contains interactions. This is an important
ingredient required in a generalized description of the plasma dynamics
in a spatially varying background.
To interpret our solutions we study the properties of spectral integrals
of test functions for dynamical quantities. We argue that localization in
space implies that the usual quasiparticle picture has to be extended to
include a limited amount of correlations (quantum coherence information)
in momentum space.

Before considering the semiclassical method for the propagator, we
derive the WKB dispersion relation from the wave equation. For simplicity we
consider a stationary case, such that $m^2\! =\! m^2(x)\,$ with $x$ defined
by $x^\mu\! =\! (0,x,0,0)$. Ignoring ${\cal L}_{\rm int}$, Eq.\
(\ref{Lagrange}) then implies
\be{wkbEOM}
(\omega^2-\vec k_{\scriptscriptstyle\parallel}^{\,2}+\partial_x^2-m^2(x))
\Phi_{\omega,\vec k_{\scriptscriptstyle\parallel}} = 0,
\ee
where $\omega$ is the conserved energy and $\vec k_{\scriptscriptstyle
\parallel}$ is the conserved momentum in the $yz$-direction. Rewriting the 
wave function as $\Phi_{\omega,\vec k_{\scriptscriptstyle\parallel}}
= u\exp{i\int_x^\infty k_x(\omega,\vec k_{\scriptscriptstyle\parallel };
x^\prime)dx^\prime}$, Eq.~(\ref{wkbEOM}) can be recast as an equation 
for the momentum variable
\be{wkb k}
k_x^2=k_0^2
-\frac{1}{2}\left(\frac{k_x^\prime}{k_x}\right)^\prime
+\frac{1}{4}\left(\frac{k_x^\prime}{k_x}\right)^2\,.
\ee
Here we used $u^\prime/u=-k_x^\prime/2 k_x$ and $k_0^2 = \omega^2-
\vec k_{\scriptscriptstyle\parallel }^{\,2}-m^2$ and {\it e.g.\/}
$k_x^{\prime}\equiv \partial_x k_x$. Eq.\ (\ref{wkb k}) can be solved
iteratively in the derivative expansion; neglecting for simplicity
$\partial_x^n m^2$ for $n\ge 2$, one obtains
\be{wkb k1}
k_x=k_0+\frac{5}{8}\frac{\delta^3}{k_0^5}
-\frac{1105}{128}\frac{\delta^6}{k_0^{11}}+{\cal O}(\delta^9) \,,
\ee
where $\delta^3\!=\!(m^{2\,\prime})^2/4$. It is known that this
constitutes a semiconvergent asymptotic series solution to (\ref{wkb k}).

\section{Semiclassical propagator}

In order to study the effect of a space-time varying mass $m^2$ on
the propagator, we use the Schwinger-Dyson equations in the Keldysh
closed time contour (CTC) formalism \cite{KBDR,Henning}. This implies
the following formally exact propagator equation in the Wigner
representation
\be{pe}
   \cos\Diamond\{k^2-m^2\pm i\omega\Gamma\}\{G^{r,a}\}=1\,.
\ee
where for simplicity we neglected the self-energy contribution. The
$\Diamond$-operator denotes a generalized Poisson bracket
$\Diamond\{f\}\{g\}\! \equiv\! (\partial_x f\partial_k g -\partial_k
f\partial_x g)/2$ and $k^\mu\! =\! (\omega,\vec k)$ is the canonical momentum
and $x^\mu=(t,\vec x)$ is from now on the average coordinate. The propagator
equation in the Wigner representation (\ref{pe}) is useful when the background
is slowly varying, \ie formally when $\partial_x\partial_k G$ is much smaller
than $G$. In our notation $G^{r,a}$
indicates the retarded and advanced propagators, whose damping terms
are $\pm i\omega \Gamma$, respectively. To the 0th order in gradients
the solution to Eq.~(\ref{pe}) reads
\be{pe0}
G^{r,a}_0 = \frac{1}{k^2-m^2\pm i\omega\Gamma}\,.
\ee
From Eq.~(\ref{pe}) we then infer that the gradient correction occurs
first at the second order \cite{Henning}.

The propagator equation (\ref{pe}) defines the spectrum of excitations in
the system. This information is encoded in the {\it spectral function}
${\cal A}=\frac{i}{2}\left(G^r-G^a\right)$ which, as a consequence of
the general equal time commutation relations, satisfies the well known
{\it sum rule}:
\be{sr}
\frac{1}{\pi}\int_{-\infty}^\infty d\omega \omega {\cal A}=1\,.
\ee
Using the complex properties of the lowest order propagator (\ref{pe0}),
one can rewrite the sum rule as a contour integral along the path shown
in Fig.\ 1 of the complex valued propagator $G_0=(\omega^2-\omega_0^2)^{-1}$,
where $\omega_0^2=\vec k^2+m^2$ and we set $\Gamma \rightarrow 0$. It is then
easy to see that  $(i/\pi)\oint_{{\cal C}_\omega} d\omega \omega G_0
\equiv \sum_{\omega=\pm\omega_0}{\rm Res}[2\omega G_0]=1$. It is convenient
to introduce a new variable $z=k^2-m^2$. In terms of $z$ the sum rule for
$G_0=1/z$ becomes simply
\be{srz}
\frac{i}{2\pi}\oint_{{\cal C}_z} dzG_0 =
  {\rm Res}[G_0]_{z=0}=1\,,
\ee
because one can deform the contour ${\cal C}_z$ to ${\cal C}_\epsilon$ around
the origin, as shown in Fig.\ 2.
Note that the
mapping $\omega\rightarrow z=\omega^2-\omega_0^2$ induces a double covering
of the complex $z$-plane; one for particles ($\Re[\omega]>0$),
and one for antiparticles ($\Re[\omega]<0$); $\Re[\omega]=0$
is mapped onto the negative real $z$-axis. Strictly speaking
the two covers are identical only when the propagator conserves CP symmetry.

\begin{figure}[t]
\centering
\hspace*{-2mm}
\leavevmode\epsfxsize=5cm \epsfbox{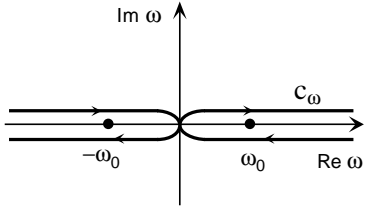}\\[3mm]
\caption[fig1]{\label{fig1} The integration contour for spectral
integral in complex $\omega$-plane.}
\end{figure}

\begin{figure}[t]
\centering
\hspace*{0mm}
\leavevmode\epsfxsize=5cm \epsfbox{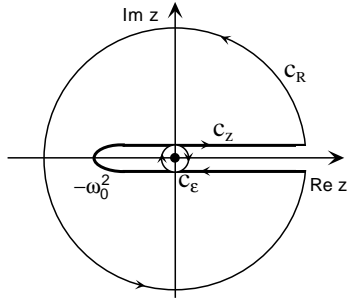}\\[3mm]
\caption[fig2]{\label{fig2} The integration contours for spectral
integral in the complex $z$-plane.  }
\end{figure}

The sum rule can also be computed by integrating along a circle ${\cal C}_R$
at $R\! \rightarrow \! \infty$, shown also in figure 2.  We now assume that
this is true in general for the full physical propagator $G$. This means that
$G$ must have a residue equal to unity at $|z|\! = \! \infty$, which is
equivalent to the following simple boundary condition
\be{bc}
G\rightarrow \frac{1}{z}\,,\qquad {\rm when}\; |z|\rightarrow\infty .
\ee
More generally we will assume that any spectral integral of the form
$\int d \omega \omega {\cal A}{\cal T}$, where ${\cal T}$ is some dynamical
quantity, is to be computed by a contour integral of $G{\cal T}$ over a contour
${\cal C}$ which contains contributions from all poles and possible
discontinuities of the propagator.

As was already pointed out, the first nontrivial correction to the propagator
equation~(\ref{pe}) occurs at the second order in gradients. To this order one
finds the following equation for the complex valued $G$:
\be{pe2}
(k^2-m^2)G
+\frac{1}{4}\left(\partial_x^2
+\frac{1}{2} m^{2\,\prime\prime}\partial_{k_x}^2\right)G = 1,
\ee
where for simplicity we considered a stationary case $m^2=m^2(x)$, again
with $x^\mu = (0,x,0,0)$. Assuming that the derivative corrections
are small, it is appropriate to use an iterative procedure, which to the
lowest nontrivial order gives
\be{G2}
G_2=\frac{1}{z}-\frac{\delta_2}{z^3}-\frac{\delta_1}{z^4}
\ee
where $\delta_2=m^{2\,\prime\prime}/2$, and
$\delta_1=(m^{2\,\prime})^2/2+k_z^2 m^{2\,\prime\prime}$. A similar
iterative procedure was used in Eq.~(\ref{wkb k1}), when constructing the
WKB approximation to the wave function. In the rest of this letter we explain
in what sense this iterative procedure can be interpreted as the semiclassical
approximation to the propagator.

Let us mention that the multiple pole form of the propagator~(\ref{G2}) can,
in a loose sense, be reconciled with the picture of quasiparticle excitations.
To see this, note that to the second order in gradients $G_2$ can equivalently
be written as
\be{G2b}
G_2^\prime=\frac{z^2}{z^3+\delta_2 z+\delta_1}
\equiv \sum_{n=0}^2\frac{a_n}{z-z_n}\,,
\ee
where $a_n$ denote the residues at the simple poles $z_n$. These residues
and poles are in general complex however, yielding unphysical complex 
dispersion
relations $z=z_n$ and weights $a_n$.  Indeed, the true physical meaning of the
semiclassical propagator emerges only when $G$ is understood in an operational
sense inside the spectral integral over the complex path in the sense
conjectured above.

\section{Recovering the WKB result}

To study the physical consequences of Eq.~(\ref{pe}) in more detail,
without overly complicating the calculations, we now make the same simplifying
assumption that was used to obtain the equation (\ref{wkbEOM}): $\partial_x^n
m^2=0$ for $n \ge 2$. Without further approximations Eq.~(\ref{pe}) 
then becomes
\be{Airy}
\delta^3\frac{d^2G}{dz^2} + zG= 1 \,,
\ee
where $\delta_1/2\rightarrow \delta^3=(m^{2\,\prime})^2/4$ and $z=k^2-m^2$.
The standard solution of (\ref{Airy}) is known. The associated Airy function
$\rm Hi(z)$  \cite{AbSte} is unphysical however, because it does not 
satisfy the
boundary condition~(\ref{bc}) and it has no isolated poles, making it 
impossible
to satisfy the sum rule. However, we can obtain a formal series solution with
the desired properties through an iterative procedure:
\be{GN}
G = \lim_{N\rightarrow \infty}
\frac{1}{z}\sum_{n=0}^N(-1)^n \frac{(3n)!}{3^{n}n!}
\left(\frac{\delta}{z}\right)^{3n} \,.
\ee
This can further be formally resummed to yield
\be{GLambda}
G = \lim_{\Lambda\rightarrow \infty}\frac{1}{z}
\int_0^\Lambda du e^{-u-\frac{1}{3}(u\delta/z)^{3}}
\,.
\ee
In the limit $\Lambda\rightarrow \infty$, $G$ is in fact equivalent
to the associated Airy function for positive real $z$, but it diverges
for $\Re[z^3]< 0$. This divergent behaviour is to be expected due to the
unbounded nature of Eq.~(\ref{Airy}), and it is also present in the series
solution, which is formally divergent for any finite $|z|$. However,
we show below that taking a finite $\Lambda$ in (\ref{GLambda}), or
truncating the {\em asymptotic} series in Eq.\ (\ref{GN}) to some
finite $N$, leads to a meaningful regularization, and in fact to the
best estimator for the propagator.

To this end consider spectral integrals of the form
\be{GT}
   G[{\cal T}] \equiv \frac{2}{\pi}\int_0^\infty dk_x \calA {\cal T}
   \rightarrow
     \frac{i}{2\pi}\oint_{{\cal C}_z}\frac{dz}{\sqrt{k_0^2-z}}G{\cal T},
\ee
where $z=k_0^2-k_x^2$ with $k_0^2 = \omega^2-\vec k_{\scriptscriptstyle
\parallel }^{\,2}-m^2$, and the contour ${\cal C}_z$ is the one shown in
figure 2. The test function $\cal T$ representing a dynamical quantity
can be any meromorphic function with no poles on the real $\omega$-axis.
Note that, as a consequence of separability of spectral and dynamical
properties, this is true for the important example of the generalized
distribution function in the dynamical Schwinger-Dyson equation.
Furthermore, ${\cal T}$ must admit a Taylor expansion around the
propagator poles, \ie it admits gradient expansion.

We chose integration over momenta rather than frequencies in Eq.\ (\ref{GT})
to facilitate direct comparison with the standard form of the semiclassical
result in Eq.~(\ref{wkb k1}). Expanding ${\cal T}$ into the Taylor series in
$k$ around $k=k_0$, and computing the residue one obtains
\be{GTt}
G[{\cal T}] = {\rm Res}\left[G{\cal T}/\sqrt{k_0^2-z}\,\right]_{z=0}
             = \sum_{n=0}^\infty t_n {\cal T}^{(n)}_0\,
\ee
where ${\cal T}_0\equiv {\cal T}(k_0)$,
${\cal T}_0^\prime\equiv (\partial_{k_x}{\cal T})_{k_x=k_0}$, {\it etc.\/}
It suffices to consider the first two terms in (\ref{GTt}) to obtain the
semiclassical dispersion relation.  Indeed,
\bea{GTJ}
G[{\cal T}] &=& \frac{{\cal T}_0}{k_{\rm sc}}
+ \left(1-\frac{k_0}{k_{\rm sc}}\right){\cal T}_0^\prime +
                     {\cal O}(\cal T^{\prime\prime })
  \nonumber \\
  &=& \frac{{\cal T}(k_{\rm sc})}{k_{\rm sc}} +
                     {\cal O}(\cal T^{\prime\prime })
\,,
\eea
where
\be{J}
\frac{k_0}{k_{\rm sc}}  = \frac{1}{\sqrt{\pi}}\sum_{n=0}^\infty(-1)^n
   \frac{\Gamma(3n+1/2)}{3^n n!}
   \left(\frac{\delta}{k_0^2}\right)^{3n}.
\ee
It can now be easily verified by inspection, that $k_{\rm sc}$ is equivalent
to the WKB momentum $k_x$ in Eq.~(\ref{wkb k1}). Moreover, one finds that
under the assumption $\partial_x^n m^2=0$ for $n\ge 2$, Eq.\ (\ref{wkb k})
reduces to the following differential equation: $y''' + \zeta y' + y/2 = 0$
for $y = \delta^{1/2}/k_{\rm sc}$, where $\zeta \equiv k_0^2/\delta$ and
$y' \equiv dy/d\zeta$, also satisfied by the solution (\ref{J}).

In figure 3 we plot the relative deviation $\kappa=k_{\rm sc}/k_0-1$ from
the unperturbed momentum $k_0$ {\it vs.\/} $\delta/k_0^2$ for the three lowest
order solutions (2nd, 4th and 6th order in gradients) as well as the formally
resummed integral representation
\be{Jint}
\frac{k_0}{k_{\rm sc}} = \int_0^\infty \frac{du}{\sqrt{\pi u}}
      \; e^{-u-\frac{1}{3}(u\delta/k_0^2)^3}\, .
\ee
For small $\delta/k_0^2$ the agreement is extremely good and it becomes
progressively worse as $\delta/k_0^2$ approaches 0.3.  It can be shown that
the highest possible accuracy of the order $\exp -(2/3)(k_0^2/\delta)^{3/2}$
is reached by the propagator $G_{2N}$ truncated at the $2N$-th order in
derivatives, where $N$ is defined by $\delta \approx \Delta_{2N} = k_0^2
(1/3N)^{2/3}$.  In fact the integral expression is just the
envelope of $G_{2N}$'s  at $\delta\approx \Delta_{2N}$, and hence represents
the best possible estimate of $k_{\rm sc}$ for any value $\delta\le \Delta_2$.
However, even though the integral gives a reasonable result also for large
values $\delta>\Delta_2$, one should be careful in attributing a physical
meaning to it, because analytic continuation of an asymptotic series is not
unique.

It is interesting to note that in the case of the electroweak phase 
transition the
expanding phase transition front typically has a width $\sim 
10-20/T$\cite{revw},
so that typically $\delta/k_0^2 \lsim 0.1$. In this region the 
asymptotic series
is accurate to 10 decimal points.

\begin{figure}[t]
\centering
%For Mac-picture:
%\hspace*{-3mm}
%\hspace*{0mm}
%\leavevmode\epsfysize=8cm \epsfbox{GFig3.eps}\\[3mm]
\vspace*{-2mm}
\hspace*{-3mm}
\leavevmode\epsfysize=8.0cm \epsfbox{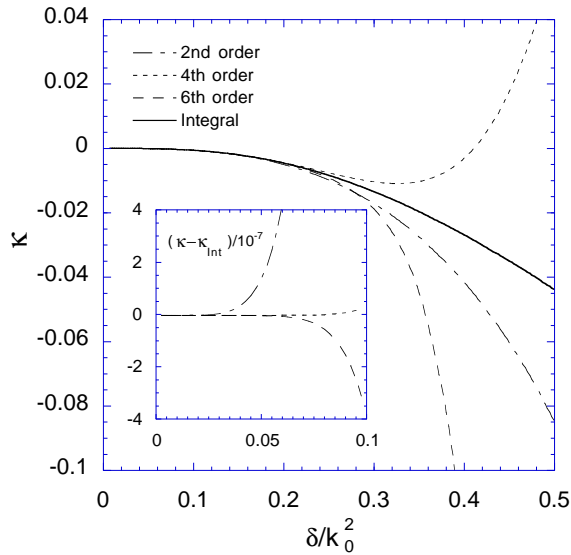}\\[-3mm]
\caption[fig3]{\label{fig3} Plotted is the quantity $\kappa \equiv
k_{\rm sc}/k_0 - 1$ for three lowest order gradient approximations for
$G$ along with the formal integral solution (solid line).  Inset: deviation
of $\kappa_n$ from the integral result for small $\delta/k_0^2$.  }
\end{figure}

We have seen that the semiclassical spectral function projects a dynamical
quantity $\cal T$ ``on-shell'' by shifting the energy hypersurface from $k_0$
to $k_{\rm sc}\!=\! k_{\rm sc}(\omega, \vec 
k_{\scriptscriptstyle\parallel };x)$.
This projection however induces additional structure in  Eq.~(\ref{GTt})
contained in the higher order derivative terms ${\cal T}^{\prime\prime}_0$,
{\it etc.\/}, and hence the semiclassical limit in field theory cannot be 
described by a single on-shell function and the semiclassical dispersion
relation. This is a direct consequence of the higher order poles in
the semiclassical propagator, which occur generically in the presence of
a space-time varying background. Note that
only a {\it finite} number of terms in the expansion in ${\cal T}^{(n)}_0$ in
Eq.~(\ref{GT}) contribute. Indeed, the coefficient of the $n$-th order
derivative ${\cal T}^{(n)}_0/n!$ has the form
$(\sqrt{k_0^2-z}-k_0)^n \! =\! (-1)^n(z/2k_0)^n \! +\! {\cal O}(z^{n+1})$,
and hence the asymptotic series representation (\ref{GN}) of $G$
implies that, at the order $2N$ in gradients, only the elements
${\cal T}^{(n)}_0$ with $n \le 3N$ can have nonvanishing coefficients.
For example, using the second order approximation to the propagator
in Eqs.~(\ref{G2}--\ref{G2b}), the series in Eq.~(\ref{GTt}) has only 4
terms, terminating with $(\delta^3/24k_0^4){\cal T}_0^{\prime\prime\prime}$.
The meaning of these corrections can be fully appreciated only when
studying the semiclassical approximation of concrete dynamical quantities.
Of particular importance is the quantum Boltzmann equation, which can be
obtained from the dynamical Schwinger-Dyson equation by the method
of spectral projection described in this letter.  In this case the test
function is related to the density matrix of the system and so the higher
order derivative terms represent nonvanishing off-diagonal correlations
between different momentum states. This is just a reflection of the
uncertainty principle, according to which localisation in the position
necessarily implies delocalization in the momentum.

\section{Conclusions}

In this letter we studied the semiclassical limit in field theory.
We found that in the presence of slowly varying backgrounds, there is a simple
iterative procedure by which one can construct the semiclassical propagator
as a semi-convergent asymptotic series controlled by powers of gradients.
While we discussed at length the simple case $\partial_x^n m^2 = 0$ for
$n\ge 2$, the method can straightforwardly be extended to the general case
with higher derivatives and nonvanishing self-energy. The semiclassical
dispersion relation appears in a nontrivial way through study of spectral
integrals and cannot be associated to a simple shift of the pole of the
propagator. We further found that the spectral projection of a dynamical
quantity necessarily contains higher order momentum derivatives as a
reflection of localization in coordinate space. As a consequence the plasma 
dynamics is not described by a single classical Boltzmann equation, but 
instead by a set of independent equations describing not only evolution
of the distribution function, but also of a limited number of its
derivatives. We will pursue this issue in a forthcoming publication.

\section*{Acknowledgements}

We thank Robert Brandenberger, Dietrich B\"odeker and Kari Rummukainen for
many discussions and valuable comments on the manuscript.
%
%
%%%%%%%%%%%%%%%%%%%%%%%%%%%%%%%%%%%%%%%%%%%%%%%%%%%%%%%%%%%%%%%%%%%%%%%%%%%%%%%
%  REFERENCES
%%%%%%%%%%%%%%%%%%%%%%%%%%%%%%%%%%%%%%%%%%%%%%%%%%%%%%%%%%%%%%%%%%%%%%%%%%%%%%%
%
% These are "Nuclear Physics"-type macrros
% ----------------------------------------------------------------------------
%
\nc{\AP}[3]    {{\it Ann.\ Phys.\ }{{\bf #1}, {#2} {(#3)}}}
\nc{\PL}[3]    {{\it Phys.\ Lett.\ } {{\bf #1}, {#2} {(#3)}}}
\nc{\PR}[3]    {{\it Phys.\ Rev.\ } {{\bf #1}, {#2} {(#3)}}}
\nc{\PRL}[3]   {{\it Phys.\ Rev.\ Lett.\ } {{\bf #1}, {#2} {(#3)}}}
\nc{\PREP}[3]  {{\it Phys.\ Rep.\ }{{\bf #1}, {#2} {(#3)}}}
\nc{\RMP}[3]   {{\it Rev.\ Mod.\ Phys.\ }{{\bf #1}, {#2} {(#3)}}}
\nc{\IBID}[3]  {{\it ibid.\ }{{\bf #1}, {#2} {(#3)}}}
%
% ----------------------------------------------------------------------------
%

\end{document}